\documentclass[12pt] {iopart}
\usepackage{graphicx}
\usepackage{color}

\expandafter\let\csname equation*\endcsname\relax

\expandafter\let\csname endequation*\endcsname\relax

\usepackage[libertine]{newtxmath}
\usepackage{slashed}
\usepackage{amsfonts}

\usepackage{amsmath}
\usepackage{microtype}\usepackage[colorlinks,linkcolor=blue,anchorcolor=blue,citecolor=blue,urlcolor=blue,breaklinks=true]{hyperref}

\def\a{\alpha}

\def\g{\gamma}

\def\G{\Gamma}

\def\pl{\partial}
\def\hs{\hspace}
\def\ol{\overline}

\def\lf{\left}
\def\rg{\right}

\begin{document}

\title{Gluon EMC Effects in Nuclear Matter}

\author{X G Wang$^1$, W Bentz$^2$, I C Clo\"et$^3$, A W Thomas$^1$}
\address{$^1$ ARC Centre of Excellence for Dark Matter Particle Physics and CSSM, Department of Physics, University of Adelaide, Adelaide SA 5005, Australia}
\address{$^2$ Department of Physics, School of Science, Tokai University, 4-1-1 Kitakaname, Hiratsuka-shi, Kanagawa 259-1292, Japan}
\address{$^3$ Physics Division, Argonne National Laboratory, Argonne, IL 60439 USA}
\eads{\mailto{xuan-gong.wang@adelaide.edu.au}, \mailto{bentz@keyaki.cc.u-tokai.ac.jp}, \mailto{icloet@anl.gov}, \mailto{anthony.thomas@adelaide.edu.au}}

\begin{abstract}
We investigate the gluonic structure of nuclei within a mean-field model of nuclear structure based upon the modification of the structure of a bound nucleon, with the nucleon described by the Nambu--Jona-Lasinio model. This approach has been shown to reproduce the European Muon Collaboration (EMC) effect, involving the ratio of the spin-independent structure functions of a heavier nucleus to that of the deuteron. It also predicts a significant nuclear modification for the spin structure functions, known as the polarized EMC effect. Here we report sizeable nuclear modifications of the gluon distributions (a ``gluon EMC effect'') for the ratios of both the unpolarized and polarized gluon distributions in nuclear matter to those of a free nucleon.
\end{abstract}


\submitto{\jpg}

\maketitle


%%%%%%%%%%%
\section{Introduction}
The European Muon Collaboration (EMC) effect~\cite{EuropeanMuon:1983wih,EuropeanMuon:1988lbf,BCDMS:1987upi,Gomez:1993ri} -- which refers to the observation that the ratios of the spin-independent structure functions of heavier nuclei to that of the deuteron, as a function of Bjorken-$x$, [$F_{2A}(x)/F_{2D}(x)$], differ substantially from unity in the valence region -- is still one of the most challenging topics in modern nuclear physics. It has triggered numerous experimental and theoretical investigations into dynamical mechanisms that might potentially generate the EMC effect. However, it has now become clear that a quantitative description of the EMC effect requires a change in the internal structure of the nucleons~\cite{Geesaman:1995yd,Miller:2001tg,Benesh:2003fk,Cloet:2019mql} bound within nuclei.

Here we shall be concerned with one particular approach to nuclear structure, based upon the modification of the structure of the bound nucleons by the scalar and vector mean-fields in nuclei. The first model within which this idea was applied to nuclear structure was the quark meson coupling (QMC) model~\cite{Guichon:1987jp,Fleck:1990td,Guichon:1995ue,Guichon:2018uew}, which was used to calculate the EMC effect soon after its discovery~\cite{Thomas:1989vt}. While that initial work was based upon the MIT bag model~\cite{Chodos:1974je}, more recently, using an analogous approach, a  theory of nuclear structure based upon the Nambu-Jona-Lasinio (NJL) model~\cite{Nambu:1961fr,Nambu:1961tp} has been developed~\cite{Bentz:2001vc,Mineo:2003vc}. This model  exhibits similar properties to the QMC model, with the advantage that it is completely covariant. Having demonstrated that it yields an excellent description of both the spin-independent and spin-dependent parton  distributions of the proton~\cite{Cloet:2005pp}, the model was then applied to investigate nuclear medium modifications to structure functions, successfully reproducing the unpolarized EMC data across the periodic table~\cite{Cloet:2006bq,Cloet:2009qs,Cloet:2012td}. This framework was also extended to the spin-dependent case, predicting a significant effect in the ratio of polarized structure functions, $g_{1A}(x)/g_{1p}(x)$, for nuclei where the spin is largely carried by the protons and in nuclear matter~\cite{Cloet:2006bq,Cloet:2005rt}.

While the EMC effects associated with the quark content of nuclei have been investigated extensively, there have been very few experimental and theoretical studies of the nuclear modification of the gluon distribution or ``gluon EMC effect''.  Indeed, the only experimental indication of the behaviour of nuclear gluons available was reported recently by the CMS Collaboration. They reported that the gluon parton distribution function (PDF) at large Bjorken-$x$ in lead ions is strongly suppressed with respect to the gluon PDF in a free
nucleon~\cite{CMS:2018jpl}. On the other hand, studies of $J/\Psi$ production on Sn and C by the New Muon Collaboration~\cite{NewMuon:1991iwd} suggested that there might be an enhancement of the gluon distribution in these nuclei. Exploring such changes with unprecedented precision is one of the primary scientific goals of the planned Electron-Ion Collider (EIC)~\cite{Accardi:2012qut,NAP25171,AbdulKhalek:2021gbh}. Lastly, lattice QCD simulations of the gluonic structure of light nuclei were reported in~\cite{Winter:2017bfs,Detmold:2020snb} for large quark masses, corresponding to $m_{\pi} \sim 450$ and $806\,$MeV, finding a hint of a small EMC-like effect in the gluon momentum fractions in the spin-averaged nuclei. 

In this work we focus on the gluon EMC effects within the framework provided by the NJL model and the QCD evolution of the nucleon and nuclear PDFs. 
Medium modifications of the gluon distributions are calculated for the first time.
We make predictions for two new EMC ratios, that is, the (per-nucleon) ratios of the unpolarized and polarized gluon distributions of nuclear matter to those of the free proton, $g_{A}(x)/g_{p}(x)$ and $\Delta g_{A}(x)/ \Delta g_{p}(x)$. For the spin-dependent case we take one proton in symmetric nuclear matter to be polarized, as the proton and neutron gluon PDFs are equal.
 
%===============================================================================
%===============================================================================
\section{Calculation and PDF Results}
\label{sec:PDF}
The NJL model is an effective model based upon the chiral symmetry of QCD  
that is characterized by a 4-Fermi contact interaction of the form, 
${\cal L}_I = \sum_i G_i \,\big(\ol{\psi}\,\G_i\,\psi\big)^2$,
where the $\G_i$ are matrices in Dirac, colour and flavour space, and the $G_i$ are dimensionful
coupling constants~\cite{Nambu:1961fr,Nambu:1961tp}. Applying Fierz transformations, 
the interaction Lagrangian can be decomposed into various
interacting $q \bar{q}$ and $qq$ channels. For the present purposes we need
\begin{equation}
{\cal L} = \ol{\psi}\lf(i\!\! \not\!\pl - m\rg)\psi + {\cal L}_{I,\pi} + {\cal L}_{I,s} + {\cal L}_{I,a},
\end{equation}
with $m$ the current quark mass and the interaction terms read
\begin{align}
&\hs{-1.5mm}{\cal L}_{I,\pi} = \frac{1}{2}\, G_\pi \lf(\lf(\ol{\psi}\psi\rg)^2 - 
\lf(\ol{\psi}\,\g_5\vec{\tau}\,\psi \rg)^2\rg), \\
&\hs{-1.5mm}{\cal L}_{I,s}   = G_s \Bigl(\ol{\psi}\,\g_5 C \tau_2 \beta^A\, \ol{\psi}^T\Bigr)
\Bigl(\psi^T\,C^{-1}\g_5 \tau_2 \beta^A\, \psi\Bigr), \allowdisplaybreaks\\
&\hs{-1.5mm}{\cal L}_{I,a}   = G_a \Bigl(\ol{\psi}\,\g_\mu C \tau_i\tau_2 \beta^A\, \ol{\psi}^T\Bigr)
\Bigl(\psi^T\,C^{-1}\g^{\mu} \tau_2\tau_i \beta^A\, \psi\Bigr),
\label{eqn:Lint}
\end{align}
where $\beta^A = \sqrt{\frac{3}{2}}\,\lambda^A~(A=2,5,7)$
are the colour $\ol{3}$ matrices and $C = i\g_2\g_0$.
${\cal L}_{I,\pi}$ generates the constituent quark mass, $M$, via
the gap equation and the pion emerges as a $q \bar{q}$ bound state.   
The terms ${\cal L}_{I,s}$ and ${\cal L}_{I,a}$ 
generate the interactions in the
scalar ($J^P = 0^+, T = 0, \text{colour}\,\ol{3}$) and axial-vector 
($J^P = 1^+, T = 1, \text{colour}\,\ol{3}$) diquark 
channels. These yield diquark states from which the nucleon is constructed as a quark-diquark bound state. 
While the couplings $G_\pi$, $G_s$ and $G_a$ are related to 
the original couplings, $G_i$,
via the Fierz transformation. To regularize the NJL model we use the proper-time scheme, which simulates important aspects of confinement~\cite{Bentz:2001vc,Ebert:1996vx,Hellstern:1997nv}. The coupling constants and the ultraviolet cut-off in the proper-time scheme are fixed by fitting the pion mass and decay constant as well as the constituent quark and diquark masses.
A detailed discussion of the specific values of the model parameters is given in~\cite{Cloet:2006bq,Cloet:2005rt}.
 
We solve the appropriate Bethe-Salpeter equations to obtain                              
the standard NJL results for the diquark $t$-matrices~\cite{Ishii:1995bu,Mineo:2002bg}.  
Following~\cite{Bentz:2001vc,Cloet:2014rja} these are approximated by the forms
\begin{align}
\label{taus}
\tau_{s}(q) &=  4i \, G_s\, - \frac{i g_s} {q^2 - M_s^2}, \\
\tau_a^{\mu\nu}(q) &= 4i\, G_a\, g^{\mu\nu} - \frac{i g_a}{q^2 - M_a^2} 
\left(g^{\mu\nu} - \frac{q^{\mu}q^{\nu}}{M_a^2} \right) \, ,
\label{taua}
\end{align}
where $q$ is the 4-momentum of the diquark.
The masses of the diquarks $M_s, \, M_a$ and their
couplings to the quarks $g_s, \, g_a$ are defined as the poles and residues of 
the full diquark $t$-matrices~\cite{Cloet:2014rja}.

The nucleon (quark-diquark) scattering amplitude satisfies the Faddeev equation 
\begin{align}
T =  Z + Z\,\Pi_N\,T =  Z + T\,\Pi_N\,Z,
\label{eqn:t}
\end{align} 
where $Z$ is the quark exchange kernel and $\Pi_N$ the product of a quark propagator
and a diquark $t$-matrix. 
The quark-diquark vertex function, $\Gamma_N$, is defined by the 
behaviour of $T$ near the pole
\begin{align} 
T \stackrel{p_+ \to \varepsilon_p}{\longrightarrow} 
\frac{\Gamma_N\,\overline{\Gamma}_N}{p_+ - \varepsilon_p} \, ,
\label{pole}
\end{align}
with $p$ the total 4-momentum of the quark-diquark pair.
Here, $\varepsilon_p = \frac{M_N^2}{2p^+}$ is the light-cone energy of the nucleon with $M_N$ the nucleon mass. Substituting~\eqref{pole} into~(\ref{eqn:t}) gives the homogeneous Faddeev equations for the vertex functions  
\begin{align}
\label{eqn:f}
\G_N = Z\,\Pi_N \, \G_N, \quad \text{and} \quad
\ol{\G}_N = \ol{\G}_N\, \Pi_N \,Z.
\end{align}
We employ the static approximation, where 
the momentum dependence of the quark exchange kernel, $Z$, is neglected.
Including both scalar and axial-vector diquark channels,  in the colour singlet and isospin-$\tfrac{1}{2}$ 
channel $Z$ becomes:
\begin{align}
Z = \frac{3}{M} \begin{pmatrix} 1 & \sqrt{3}\g_{\mu'}\g_5 \\
             \sqrt{3}\g_5\g^{\mu} & -\g_{\mu'}\g^{\mu} \end{pmatrix}.
\end{align}
The quantity $\Pi_N$ describes the quark-diquark bubble graph:
\begin{align}
\Pi_N(p) &= \int \frac{d^4k}{(2\pi)^4}\, \tau(p-k)\, S(k),
\end{align}
where
\begin{align}
\tau(q) =  \begin{pmatrix} \tau_s(q) & 0 \\ 0 & \tau_a^{\mu\nu}(q) \end{pmatrix},
\label{tau}
\end{align}
and $S(k)$ is the propagator for the dressed quark (mass $M$). The eigenfunction of the kernel $K \equiv Z\,\Pi_N$, in~(\ref{eqn:f}), has the following form, up to an overall normalization:
\begin{align}
\G(p,s) = \begin{bmatrix} \a_1 \\ \a_2\,\frac{p^{\mu}}{M_N}\,\g_5\ + \a_3\,\g^\mu\g_5 \end{bmatrix}u_N(p,s) \, .
\label{eqn:nvertex}
\end{align}
Here $s$ is the nucleon spin, the $\a_i$ are constants obtained by solving~(\ref{eqn:f}), the upper and lower components refer to the scalar and axial-vector diquark channels, respectively,
and $u_N$ is a free nucleon Dirac spinor with mass $M_N$. The normalization is given by the residue at the $t$-matrix pole defined by~\eqref{pole} and we dropped the isospin structure for clarity.  
Inserting this form into~(\ref{eqn:f}) gives three homogeneous equations for the $\alpha$'s, while the nucleon mass, $M_N$, is determined by the requirement that the eigenvalue of $K$, in~(\ref{eqn:f}), should be unity. 

In nuclear matter the quark mass is modified by the mean scalar field. This, in turn, leads to changes in the diquark masses 
as well as the quark spinors and this finally modifies the structure of the bound nucleon. Because the coupling of the composite nucleon to the scalar field depends on the structure of the nucleon, one must solve all of these equations self-consistently. In addition, the mean vector field shifts the energy of the quarks in medium, which results in a scale transformation of the nuclear PDFs~\cite{Mineo:2003vc}. Full details of this procedure may be found in~\cite{Mineo:2003vc,Cloet:2006bq}. 

The spin-independent quark light-cone momentum distribution in a free nucleon is defined by~\cite{Jaffe:1985je,Barone:2001sp}
\begin{equation}
\label{eq:unpol-pdf}
q(x) = -i \int\frac{d^4k}{(2\pi)^4} \delta\!\left(x-\frac{k^+}{p^+}\right) \text{Tr}\left(\gamma^+\,M(p,k)\right), 
\end{equation}
where $x$ is the Bjorken scaling variable, as usual the light-cone coordinates are defined as $a^\pm = a^0+a^3$  and $M(p,k)$ is the quark (momentum $k$) two-point function in the nucleon (momentum $p$). The spin-dependent distribution, $\Delta q(x)$, is defined by the replacement $\gamma^+\rightarrow \gamma^+\gamma_5$ in~(\ref{eq:unpol-pdf}). An analogous expression holds for the nucleon PDFs in nuclear matter, where the dressed quark propagators include effects from the mean scalar and vector fields, and Fermi motion is included following the standard convolution formalism~\cite{Cloet:2005rt,Jaffe:1985je}.

Since the NJL model has no dynamical gluons, the gluon PDFs are zero at the model scale. In addition, at the level of approximation used herein, the sea-quark PDFs also vanish at the model scale. A more sophisticated treatment including the pion would generate a non-perturbative sea distribution with $\bar{d} > \bar{u}$~\cite{Thomas:1983fh,Salamu:2019dok}. However, in the present approach both the gluon and sea-quark PDFs are generated by the QCD evolution from the model scale to the scales associated with deep inelastic scattering. As found by many earlier studies~\cite{Novikov:1976dd}, the model scale associated with a valence dominated picture is of the order of the constituent quark mass, in our case around 400 MeV. The work of Gl\"uck, Reya and Vogt established such a procedure as a phenomenologically useful method generating realistic 
PDFs~\cite{Gluck:1977ah,Gluck:1989ze,Gluck:1991ng,Gluck:1993im,Vogt:1995xs,Gluck:1994uf}.

\begin{figure}[!h] %[tbp]
\centering\includegraphics[width=8.0cm]{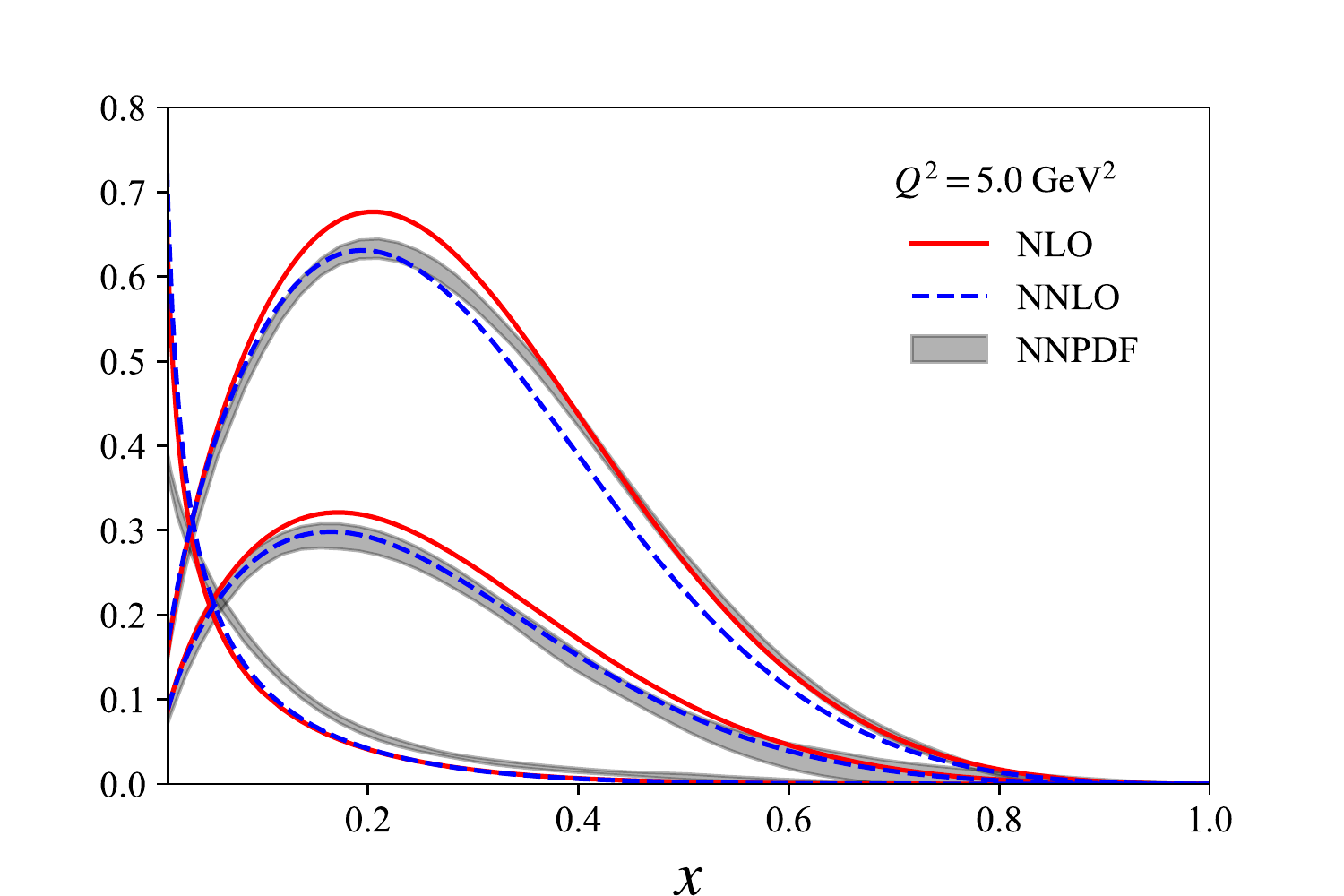}
\centering\includegraphics[width=8.0cm]{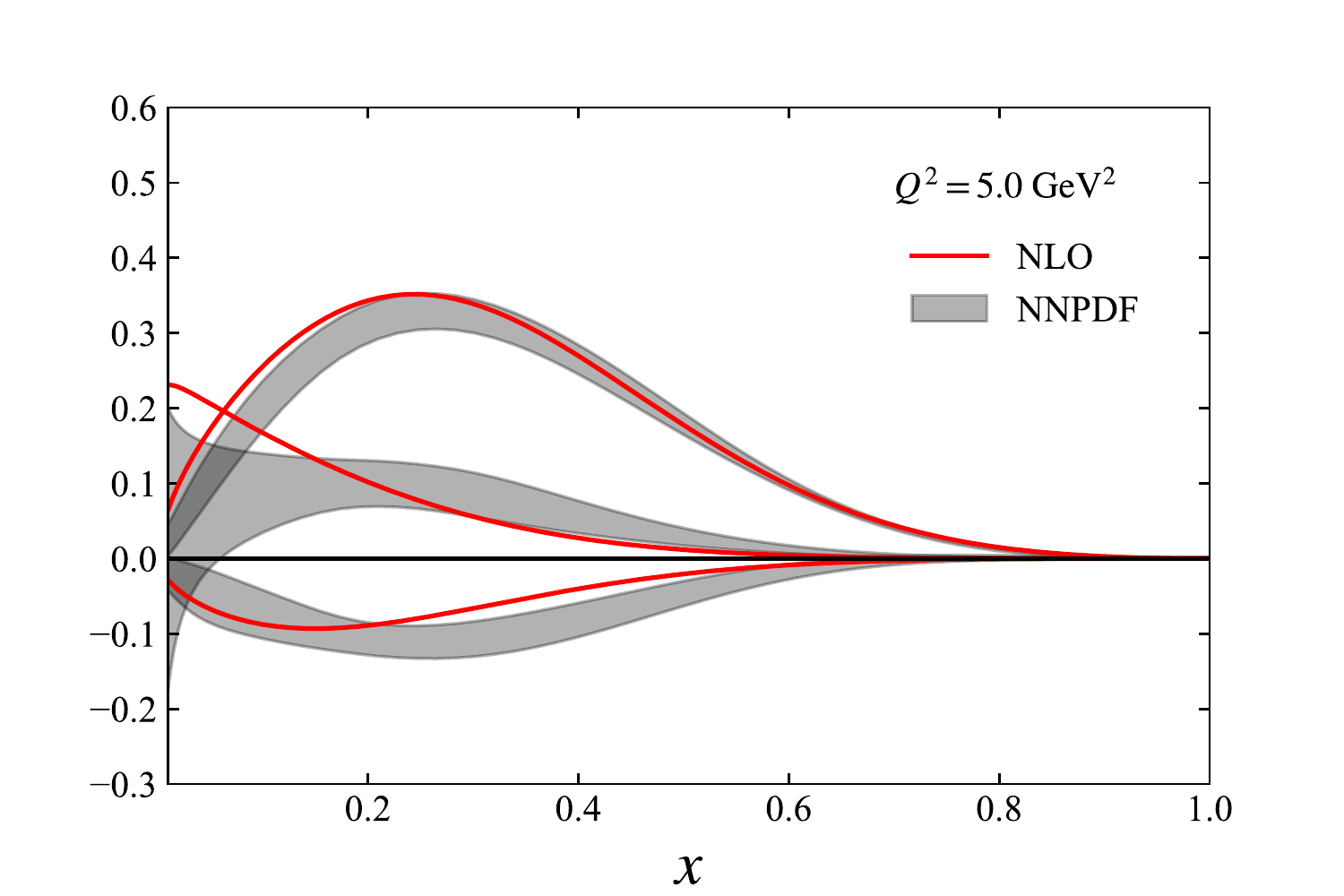}
\caption{(Colour online) Results for the spin-independent ({\it upper}) and spin-dependent ({\it lower}) parton distributions of a free nucleon obtained by QCD evolution at both NLO and NNLO to the scale $Q^2 = 5\,$GeV$^2$. From top to bottom, the groups of lines represent $x u_v$, $x d_v$, $x g/10$ in unpolarized case, and $x \Delta u_v$, $x \Delta g$, $x \Delta d_v$ in polarized case. 
The phenomenological results of unpolarized and polarized PDFs are taken from NNPDF3.0~\cite{NNPDF:2014otw} and NNPDFpol1.1~\cite{Nocera:2014gqa}, respectively.} 
\label{fig:free-PDFs}
\end{figure}

The spin-independent and spin-dependent parton distributions for a free nucleon are shown in 
Fig.~\ref{fig:free-PDFs}, both at the model scale of $Q_0^2 = 0.16\ {\rm GeV}^2$ and evolved to a scale of $Q^2 = 5\,$GeV$^2$.\footnote{We evolve our PDF results at both next-to-leading order (NLO) and NNLO from the model scale to $Q^2 = 5\,$GeV$^2$ using the APFEL program~\cite{Bertone:2013vaa}.} That the model scale is necessarily low, with $Q_0$ of order the constituent quark mass, has been recognized for many years~\cite{Jaffe:1980ti,Signal:1989yc,Diakonov:1996sr}. Because the relevant parameter for QCD evolution is $\alpha_s/4\pi$, which is less than one at this scale, it is reasonable to consider $Q_0$ a parameter of the model which is chosen so that the valence up quark distribution best reproduces the empirical parametrization after evolution at either NLO or NNLO~\cite{Cloet:2005pp}. Our results at $Q^2 = 5\,$GeV$^2$ are in good agreement with the latest phenomenological determinations from global fit analyses~\cite{NNPDF:2014otw, Nocera:2014gqa}. The level of agreement between the results obtained using NLO and NNLO evolution provide some confidence in the approach in spite of the relatively low model scale used.

\begin{figure}[!t]
\centering\includegraphics[width=8.0cm]{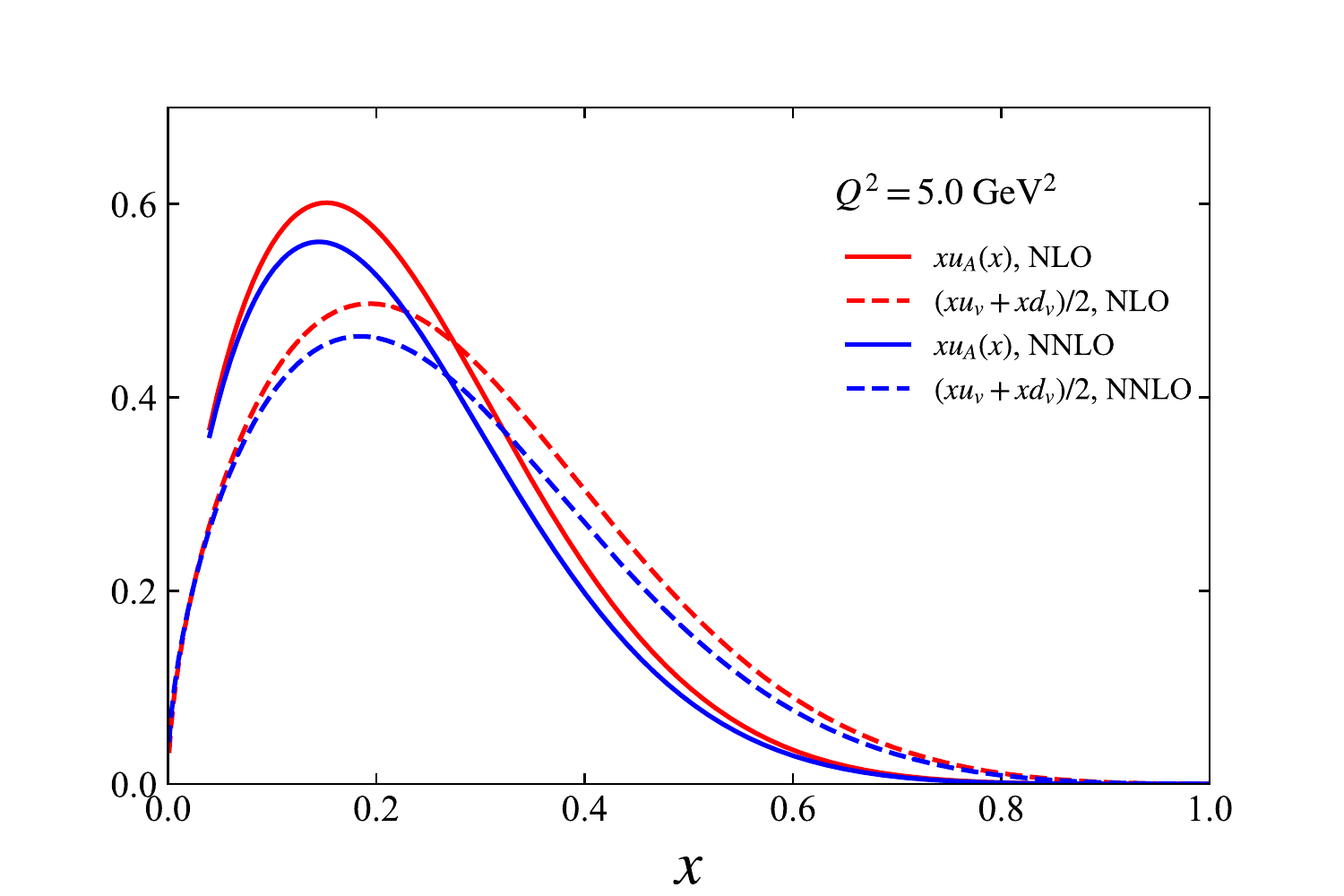}
\centering\includegraphics[width=8.0cm]{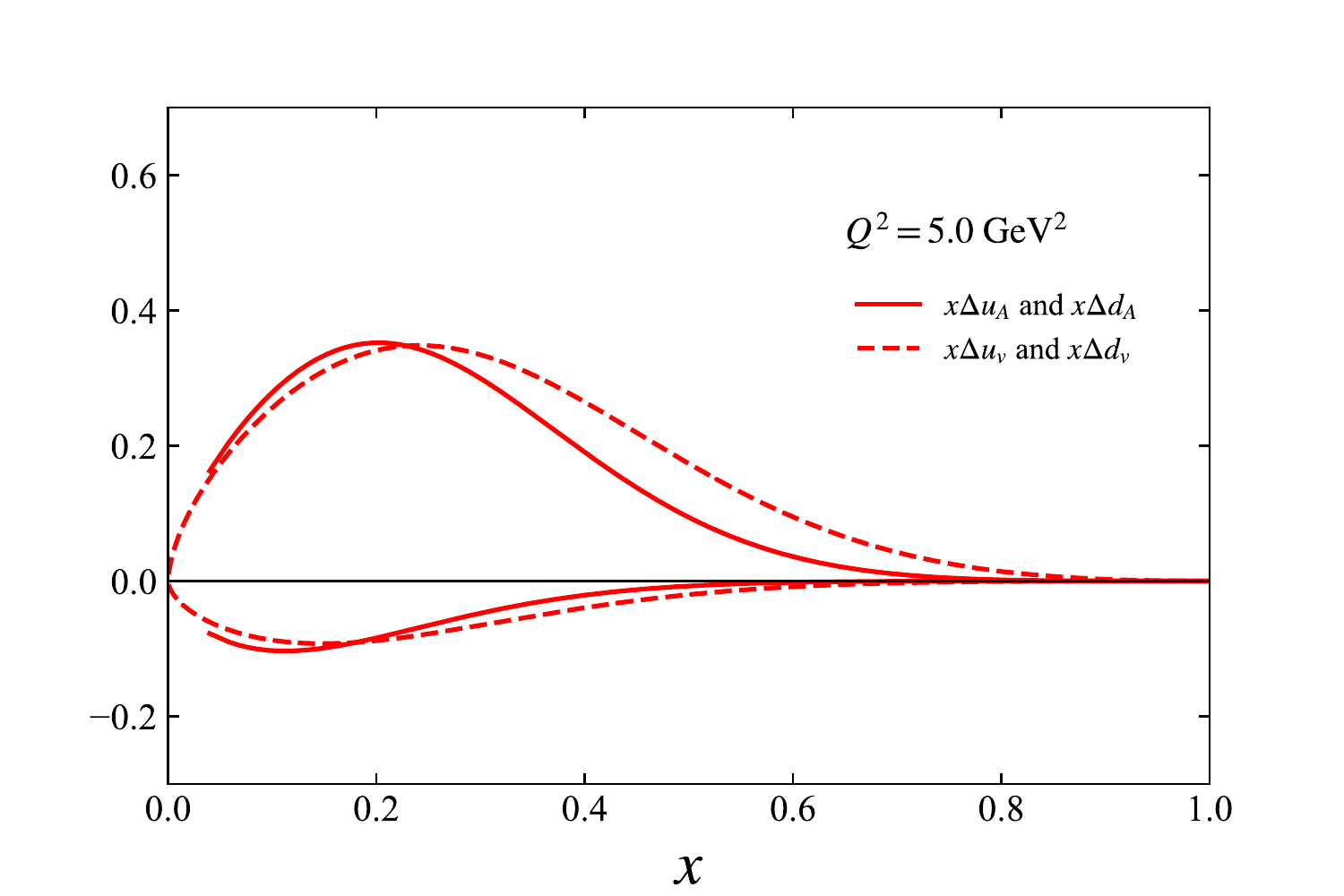}
\caption{The solid lines represent the (per-nucleon) spin-independent valence quark PDF in isospin symmetric nuclear matter ({\it top}) and the spin-dependent valence $u$ and $d$ PDF of a single polarized proton embedded in isospin symmetric nuclear matter ({\it bottom}). These results are at the scale $Q^2 = 5\,$GeV$^2$ and are compared with the free nucleon PDFs (dashed lines) at the same scale.}
\label{fig:NM-PDFs}
\end{figure}

Figure~\ref{fig:NM-PDFs} presents our results for the per-nucleon spin-independent quark distributions in isospin symmetric nuclear matter (top panel) and results for the spin-dependent quark distributions for a polarized proton embedded in isospin symmetric unpolarized nuclear matter (bottom panel) at $Q^2 = 5\,$GeV$^2$. In both cases, a sizeable reduction in the valence quark region is clearly seen.

%-------------------------------------------------------------------------------
%-------------------------------------------------------------------------------

%-------------------------------------------------------------------------------
%-------------------------------------------------------------------------------

%===============================================================================
%===============================================================================
\section{Gluon EMC Effect Results}
The unpolarized and polarized EMC effects for gluons can be defined in analogy to the more familiar EMC effects, that is:
\begin{align}
\label{eq:gemc}
R_G &= \frac{g_A(x)}{Z/A\,g_p(x) + N/A\,g_n(x)} \to \frac{g_A(x)}{g_p(x)}, \\
\label{eq:gpemc}
\Delta R_G &= \frac{\Delta g_A(x)}{P_p\,\Delta g_p(x) + P_n\,\Delta g_n(x)} \to \frac{\Delta g_A(x)}{\Delta g_p(x)},
\end{align}
where $g_p, g_n$ are the unpolarized gluon PDFs for the nucleon and $\Delta g_p, \Delta g_n$ the polarized nucleon gluon PDFs. The analogous nuclear quantities are $g_A$ and $\Delta g_A$, respectively, $Z$ is the proton number and $N$ the neutron number of the target, and $P_p$ the effective proton polarization and $P_n$ the effective neutron polarization for the nuclear target. In~\eqref{eq:gemc}--\eqref{eq:gpemc} the denominators are further simplified by assuming that the proton and neutron gluon PDFs are equal and that  $P_p + P_n \simeq 1$. The latter is a very good approximation in the most promising case for experimental investigation, namely $^7$Li.

For this initial study we consider gluon distributions in isospin symmetric ($N=Z$) nuclear matter. Because we use the NJL model, the gluon distributions vanish at the model scale, both for the free nucleon and in nuclear matter. However, compared to the free nucleon PDFs, the quark distributions in nuclear matter are modified by the scalar and vector mean fields in the nuclear medium. This medium modification changes the initial quark distributions in medium, and therefore the gluon distributions generated by QCD evolution will inherit this medium modification, and a gluon EMC effect will be generated. In a more complete calculation the gluons may be modified directly by the nuclear medium, so this initial study provides a baseline from which to study the gluon EMC effect in unpolarized and polarized PDFs.

%-------------------------------------------------------------------------------
\begin{figure}[tbp]
\centering\includegraphics[width=7.0cm]{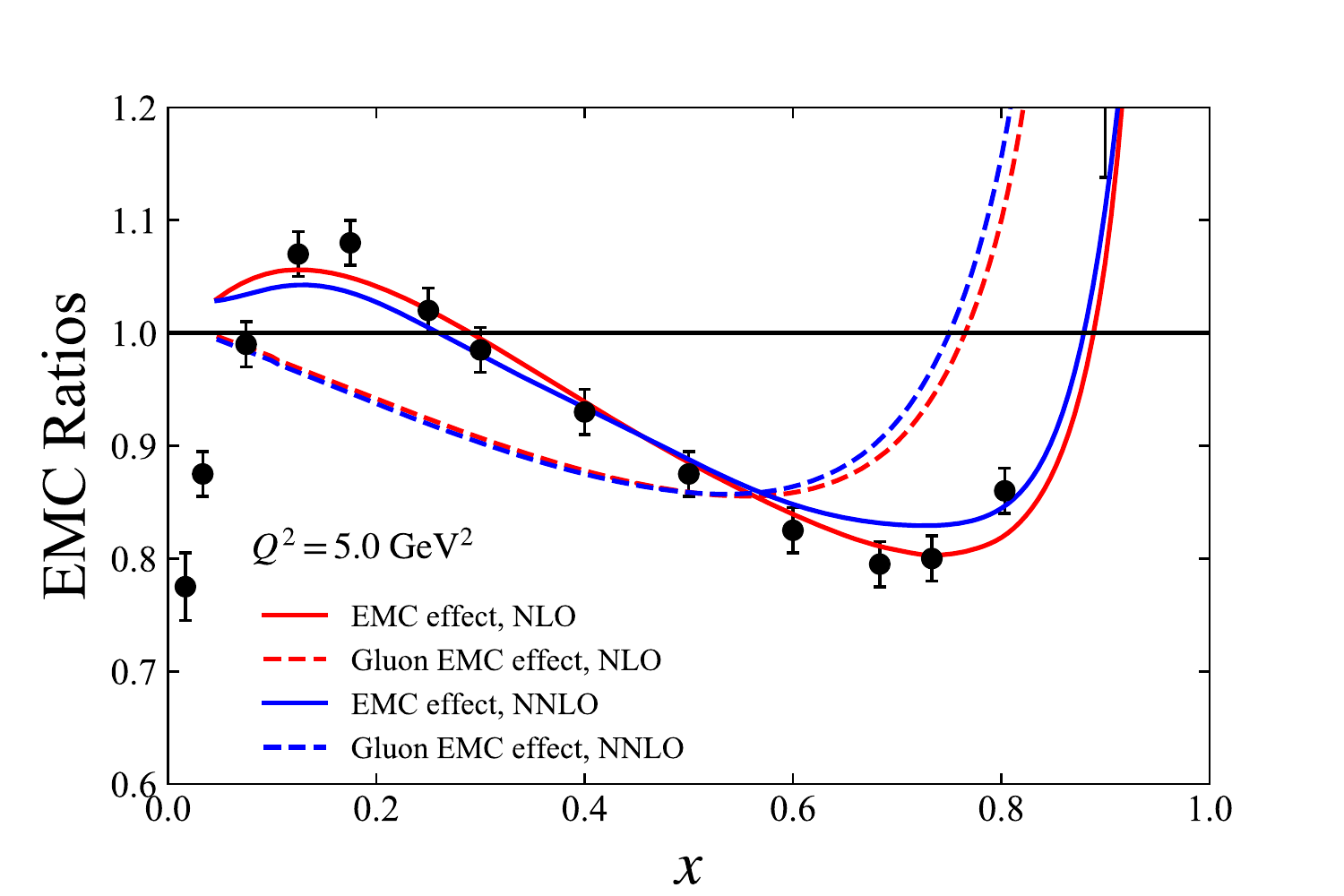} %\hfill
\centering\includegraphics[width=7.0cm]{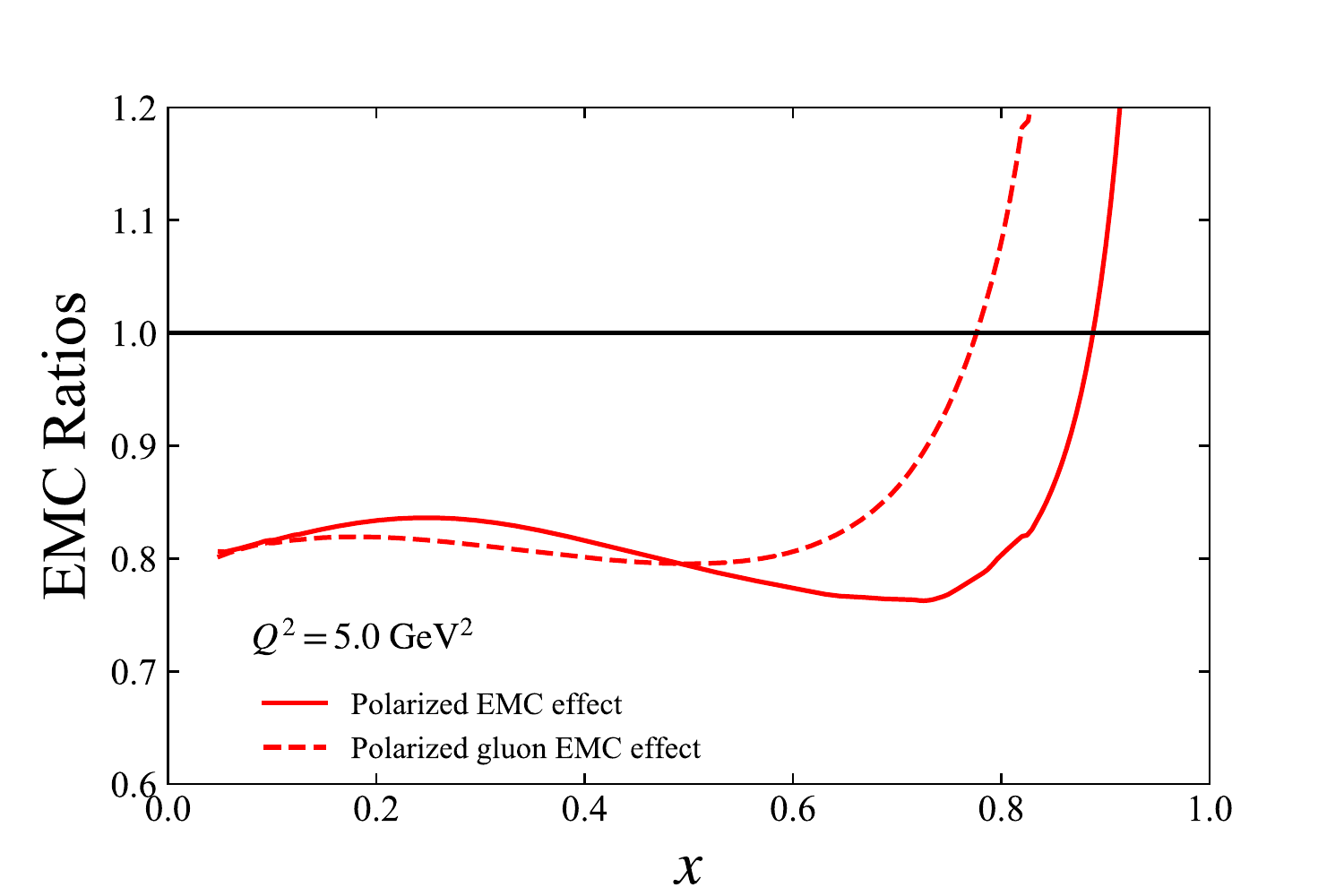}
\caption{(Colour online) ({\it Left panel}) Unpolarized EMC ratios for the structure functions $F_{2A}(x)/F_{2N}(x)$ (solid) and the unpolarized gluon distributions $g_A(x)/g_p(x)$ (dashed). ({\it Right panel}) Polarized EMC ratios for the structure functions $g_{1A}(x)/g_{1p}(x)$ (solid) and polarized gluon distributions $\Delta g_A(x)/\Delta g_p(x)$ (dashed). The empirical data points are the unpolarized nuclear matter results for the EMC ratio from Ref.~\cite{Sick:1992pw}.}
\label{fig:emc-ratios}
\end{figure}
%-------------------------------------------------------------------------------

In Fig.~\ref{fig:emc-ratios} we show our results for the EMC ratios of the unpolarized and polarized structure functions~\cite{Cloet:2006bq,Cloet:2005rt}, as well as for the gluon distributions as defined in~\eqref{eq:gemc}--\eqref{eq:gpemc}. These results are obtained from the PDF results presented in Sec.~\ref{sec:PDF}.
Because of the neglect of shadowing effects in the present calculation, the description of nuclear structure functions in the low $x$ region is incomplete, which is why this region is not shown in Fig.~\ref{fig:emc-ratios}.

The left panel in Fig.~\ref{fig:emc-ratios} shows the spin-independent EMC ratios. In the valence quark region, the model is able to reproduce the empirical nuclear matter results~\cite{Sick:1992pw} for $F_{2A}(x)/F_{2N}(x)$ at nuclear matter density extremely well~\cite{Cloet:2005rt}, where $F_{2N}(x)$ is the free isoscalar nucleon structure function and $F_{2A}(x)$ represents the in-medium per-nucleon result. Also shown in this figure is the predicted ratio of the nuclear gluon distribution (per nucleon) to that of the free nucleon as defined in~\eqref{eq:gemc}. Clearly our analysis predicts for the first time a sizeable gluon EMC effect.

The EMC effects in the spin-dependent case are given in the right panel of figure~\ref{fig:emc-ratios}. The size of the structure function ratio $g_{1A}(x)/g_{1p}(x)$ is at least as large as that of the spin-independent case, consistent with the previous results in~\cite{Cloet:2006bq,Cloet:2005rt,Smith:2005ra,Tronchin:2018mvu}. It is also clear from this figure that we find a large EMC effect for the polarized gluon distribution ratio $\Delta g_A(x)/\Delta g_p(x)$, which is again larger than the gluon EMC effect seen in the unpolarized case.

%===============================================================================
%===============================================================================
\section{Conclusion}
Starting with the NJL model to describe the structure of free nucleons, we used our previous results~\cite{Cloet:2006bq,Cloet:2005rt} to self-consistently obtain the change in the structure of nucleons bound in nuclear matter, taking into account the Lorentz scalar and vector mean-fields. These results give the spin-independent and spin-dependent quark distributions for both free nucleons and  nucleons in nuclear matter. Gluon distributions at any scale $Q^2 > Q_0^2$ were generated dynamically through both NLO and NNLO QCD evolution in order to estimate the sensitivity of the final results to the relatively low model scale. We found significant EMC effects for both the unpolarized and polarized gluon distributions. Sizeable gluon EMC effects therefore follow naturally from the observed unpolarized EMC effect and predictions for the polarized EMC effect~\cite{Cloet:2006bq,Cloet:2005rt,Smith:2005ra,Tronchin:2018mvu}. 

An experiment at Jefferson Lab is planned to measure the polarized EMC effect in $^7$Li~\cite{pemc}. The gluonic aspects of the structure of nucleons and nuclei will be accessible at the future Electron-Ion Collider (EIC)~\cite{Accardi:2012qut,NAP25171,AbdulKhalek:2021gbh}. Experimental confirmation of our predictions for the gluon EMC effects would provide further insight into QCD effects in nuclei.

\ack{
This work was supported by the University of Adelaide and the Australian Research Council through the Centre of Excellence for Dark Matter Particle Physics (CE200100008) and Discovery Project DP180100497. IC is supported by the U.S.~Department of Energy, Office of Science, Office of Nuclear Physics, contract no.~DE-AC02-06CH11357.
}

\end{document}